\documentclass{elsart}
\usepackage{graphicx}

\begin{document}

\begin{frontmatter}

\title{Power law for the calm-time interval of price changes}
\author[label1,label2]{Taisei Kaizoji\corauthref{col1}} 
\author[label2]{Michiyo Kaizoji}
\address[label1]{Division of Social Sciences, International Christian University, Mitaka, Tokyo 181-8585 Japan.}
\address[label2]{Econophysics Laboratory, 5-9-7-B Higashi-cho, Koganei-shi, Tokyo 184-0011 Japan.}
\corauth[col1]{Corresponding author: Taisei Kaizoji, E-mail: 
kaizoji@icu.ac.jp; 
URL: http://subsite.icu.ac.jp/people/kaizoji/econophysics/ (T. Kaizoji).}

\begin{abstract}
In this paper, we describe a newly discovered statistical property of time series data for daily price changes. We conducted quantitative investigation of the {\it calm-time intervals} of price changes for 800 companies listed in the Tokyo Stock Exchange, and for the Nikkei 225 index over a 27-year period from January 4, 1975 to December 28, 2001. A calm-time interval is defined as the interval between two successive price changes above a fixed threshold. We found that the calm-time interval distribution of price changes obeys a power law decay. Furthermore, we show that the power-law exponent decreases monotonically with respect to the threshold. 
\end{abstract}

\end{frontmatter}
\section{Introduction}
Certain stylized facts of financial time series are well known to be true, for example fat-tailed distribution of log returns [1], the power-law distribution of log returns [2-9], volatility clustering [10,11] which is described as on-off intermittency in the literature on nonlinear dynamics, and multi fractality of volatility [12-16]. 

Several researchers have recently investigated the statistical properties of waiting times of high-frequency financial data [17-24], and Scalas et al. [17-21] in particular have applied the theory of continuous time random walk (CTRW) to financial data. They also found that the waiting-time survival probability for high-frequency data of the 30 DJIA stocks is non-exponential [21]. 

In the present study we further examined the statistical properties of the waiting-time of stock price changes. A question of current interest in financial markets is: what are the statistical properties of the time intervals between two successive price changes? Specifically, how long after one large price fluctuation will the next large price fluctuation occur? To answer these questions, we investigated the distribution of the calm-time interval, that is, the time interval between two successive price changes above a fixed threshold [25]. We quantitatively analyzed a database covering securities of companies listed in the Tokyo Stock Exchange (TSE), as well as the Nikkei 225 index. The database covering securities of companies listed in the TSE provides daily data on closing prices and covers the 27-year period between January 4, 1975 and December 28, 2001. For the purposes of this study, we selected the 800 companies among all the companies listed that had an unbroken series of daily closing prices for the entire 27-year period. Each time series had approximately 7000 data points corresponding to the number of trading days in the 27-year period. We report here that the calm-time interval distributions of price changes are well approximated by a power-law function. Furthermore, we show that the power-law exponent decreases monotonically with respect to the threshold. 

\section{The calm-time interval distribution of price changes}
We first analyzed the selected database of stock prices. We formed 800 time series of $ p_j(t) $, which denotes the closing price of company $ j $. We defined price change $ S_j(t) $ as [an] increment of the price $ S_j(t) = p_j(t) - p_j(t-1) $ where $ p(t) $ is the closing price on trading day $ t $. We normalized price changes as follows: 
\begin{equation}
s_j(t) = \frac{[S_j(t) - \bar{S_j}]}{V_j}; \quad \bar{S_j} = \frac{1}{T}\sum_{t=1}^T S_j(t), \quad j = 1, \ldots, 800. 
\end{equation}
where $ V_j $ is the standard deviation of company $ j $ and $ \bar{S_j}(t) $ is a time average. We obtained approximately 7000 normalized price changes $ s_j(t) $ per company for the 27-year period. Using these normalized values, we first measured the calm-time intervals by introducing a threshold $ \theta $ of the absolute value of normalized price changes, $|s_j(t)|$. A calm-time interval $ \tau_j $ is defined as the time interval from a day on which the absolute value of the normalized price change is above the threshold to the next day that it again exceeds the threshold. We counted the number of calm-time intervals greater than one day, and then calculated the complementary cumulative distribution function of the calm-time intervals of the normalized price changes, $P(|\tau_j| \geq \tau)$. 

In the present study, we focused on the statistical properties of the calm-time intervals of price changes rather than those of the calm-time intervals of log returns. Although we did investigate the statistical properties of the calm-time intervals of log returns, we were unable to obtain clearly empirical results on them, while we were able to obtain empirical results on price changes. 

{\it Tomen Corporation}, a large Japanese business company based in Osaka, serves as a typical example of the 800 large companies we analyzed. Figure 1 illustrates the calm-time interval of price changes for Tomen Corp. The vertical axis shows the absolute value of the price change $ |s_j| $ and the horizontal axis indicates the date. The arrows on the figure indicate the calm-time intervals, $ \tau_{tomen} $. In this case we established a threshold of $ \theta = 0.7 $. Figure 2 presents the log-log plots of the complementary cumulative distribution of the calm-time intervals of price changes with the fixed threshold $ \theta = 0.7 $. The dots represent the observed distribution, while the solid lines represent the power-law distribution, which are expressed by 
\begin{equation}
 P(\tau_{tomen} \geq \tau) \sim \frac{1}{\tau^{\alpha}}. 
\end{equation}
A linear regression fit in the region from one to one hundred standard deviations yields 
$ \alpha = 1.13 \pm 0.02$, $ R^2 = 0.99 $. 

In order to confirm the robustness of the above analysis, we repeated the analysis for each time series of price changes for each of the 800 companies. For all companies, the asymptotic behavior of the functional form of the distribution was consistent with power laws. Figure 3, which shows the log-log plot of the complementary cumulative distributions of the calm-time intervals with a threshold of $ \theta = 0.7 $ for 10 companies randomly selected from the 800 companies, demonstrates that the calm-time intervals of the price changes obey a power-law decay. The power-law exponent $ \alpha $ is distributed in the range from 1 to 2\footnote{The estimates of the power-law exponent $ \alpha $ were found to be sensitive to the bounds of the regression used for fitting. Thus we used a coefficient of determination of $ R^2 $ of the linear regression line as a standard to determine the appropriate values of the power-law exponent $ \alpha $. We chose the results of the regression fit where the coefficient of determination was greater than $0.98$.}. Under the threshold $ \theta = 0.7 $, the estimates of the power-law exponent $ \alpha $ were within the stable L$\acute{e}$vy domain, $ 0 < \alpha \geq 2 $ for all but $4$ of the $800$ companies. Figure 4 shows the histogram of the power-law exponent $ \alpha $ obtained from regression fits to the individual complementary cumulative distribution for all $800$ companies. Fifty-seven percent of all the estimates of $ \alpha $ were between $1.3$ and $1.5$. 

We next analyzed the price series data of price series for the Nikkei 225 index in the period from January 4, 1975 to December 28, 2001. The Nikkei 225 index is the Dow-Jones average of 225 industrial stocks listed in the Tokyo Stock Exchange. Figure 5 presents the log-log plots of the complementary cumulative distribution function of the calm-time intervals for the Nikkei 225 index for the threshold $ \theta = 0.25 $. The dots represent the observed complementary cumulative distribution, and the solid line shows the power-law function, which is well expressed by $ P(\tau_{nikkei} \geq \tau) \sim \tau^{-\alpha} $. The regression fit in the region $ 1 \leq \tau \leq 90 $ yields $ \alpha = 1.13 \pm 0.02 $, $R^2 = 0.992 $. 

\subsection{Dependency of the power-law exponent on the threshold} 
Finally, we investigated the dependency of the power-law exponent $ \alpha $ on the threshold, $ \theta $. Figure 6 shows the plots of the complementary cumulative distribution functions of the calm-time intervals for Tomen Corp. under $ \theta $ ranging from $0.1$ to $0.9$, and Figure 7 shows the plot of the complementary cumulative distribution of calm-time intervals for the Nikkei 225 index under $ \theta $ ranging from $0.15$ to $0.3$. Table 1 shows the threshold $ \tau $ and the corresponding power-law exponent $ \alpha $. Note that in both cases the power-law exponent $ \alpha $ decreases monotonically with respect to the threshold $ \theta $. Our extensive examinations using price data for 800 companies provide strong evidence that this tendency is robust. 

\begin{table}[hbtp]
\begin{center}
\begin{tabular}{cc|cc} \hline
\multicolumn{2}{c|}{Tomen Corp.}& \multicolumn{2}{c}{The Nikkei 225 index}\\ \hline
\cline{1-4}
{\it threshold $ \theta $}& {\it the exponent $ \alpha $}& {\it the threshold $ \theta $}& {\it the exponent $ \alpha $} \\ \hline
0.1 & 1.81 & 0.05 & 2.47 \\
0.2 & 1.43 & 0.1 & 2.24 \\
0.3 & 1.31 & 0.15 & 1.66 \\
0.5 & 1.22 & 0.2 & 1.25 \\
0.7 & 1.13 & 0.25 & 1.17 \\
0.9 & 0.97 & 0.3 & 1.16 \\ \hline
\end{tabular}

\caption{Dependency of the power-law exponent $ \alpha $ on the threshold $ \theta $}
\end{center}
\end{table}

\section{Conclusion}
In conclusion, we have discovered a power-law scaling for calm-time intervals between large fluctuations in daily stock prices. Our results show that the power law of the calm-time intervals is very robust. The power law statistics shown here are not only an interesting theoretical finding but are presumably also a practical tool for measuring the risk of security investments. Our empirical results constitute conditions that models of financial markets would have to satisfy. In future research, we plan to model the power laws of calm-time intervals of lower-frequency financial data. Theoretical study will also be left for future work. 

\section{Acknowledgements}

The author would like to thank Enrico Scalas for his helpful comments and suggestions. Financial support from the Japan Society for the Promotion of Science under Grant-in-Aid No. 06632 is gratefully acknowledged. The authors take full responsibility for all remaining errors.

\newpage 


\begin{figure}

\begin{center}
  \includegraphics[height=20cm]{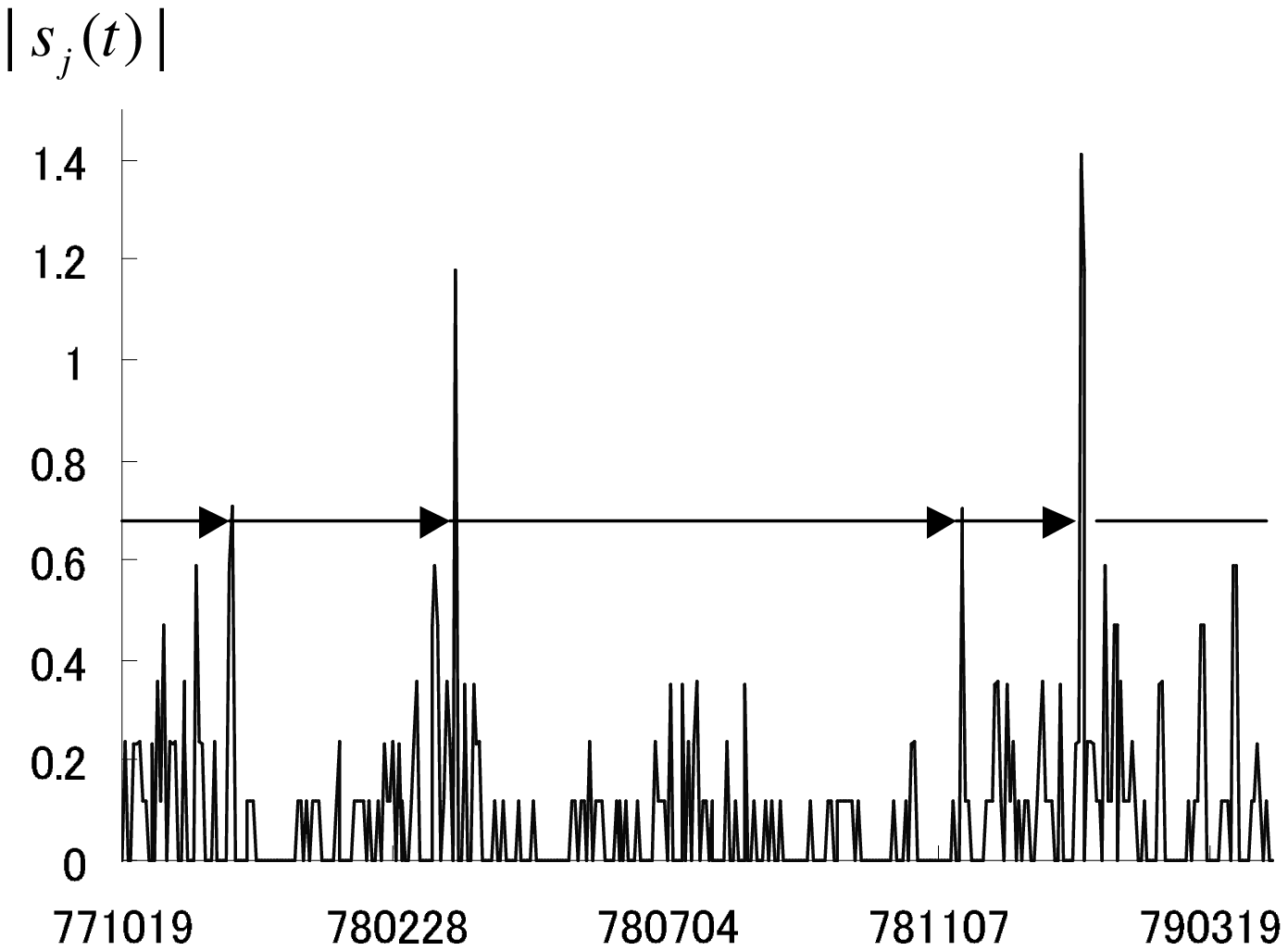}
\end{center}
\caption{The time series of the calm-time intervals of price changes for Tomen Corp. The vertical axis indicates the absolute value of the price increment $ |s_j| $, and the horizontal axis indicates the date. Arrows indicate the calm-time intervals.} 
\label{fig1}
\end{figure}
\begin{figure}
\begin{center}
  \includegraphics[height=20cm]{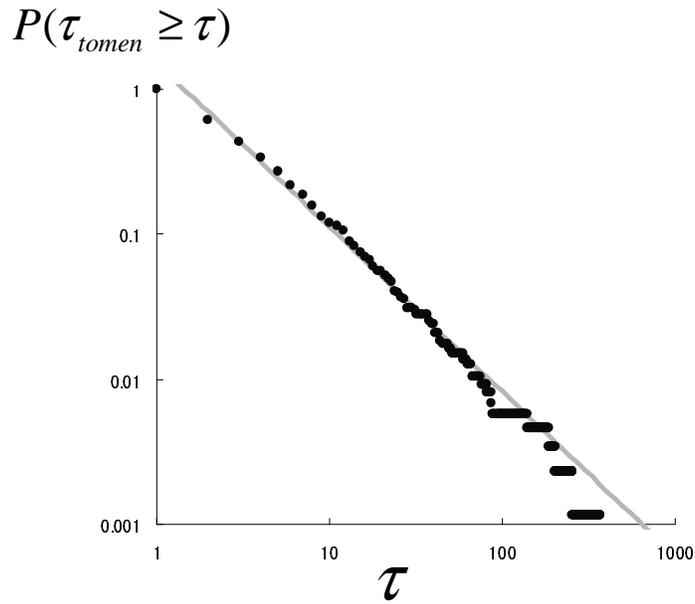}
\end{center}
\caption{The log-log plots of the complementary cumulative distribution of the calm-time intervals for Tomen Corp. in the 27-year period from January 4, 1975 to December 28, 2001 with a threshold of $ \theta = 0.7 $. Dots represent the observed complementary cumulative distribution, while solid lines indicate the power-law function, $ P(\tau_{tomen} \geq \tau) \sim \tau^{-\alpha} $ with the exponent $ \alpha = 1.13 $. } 
\label{fig2}
\end{figure}

\begin{figure}
\begin{center}
  \includegraphics[height=21cm]{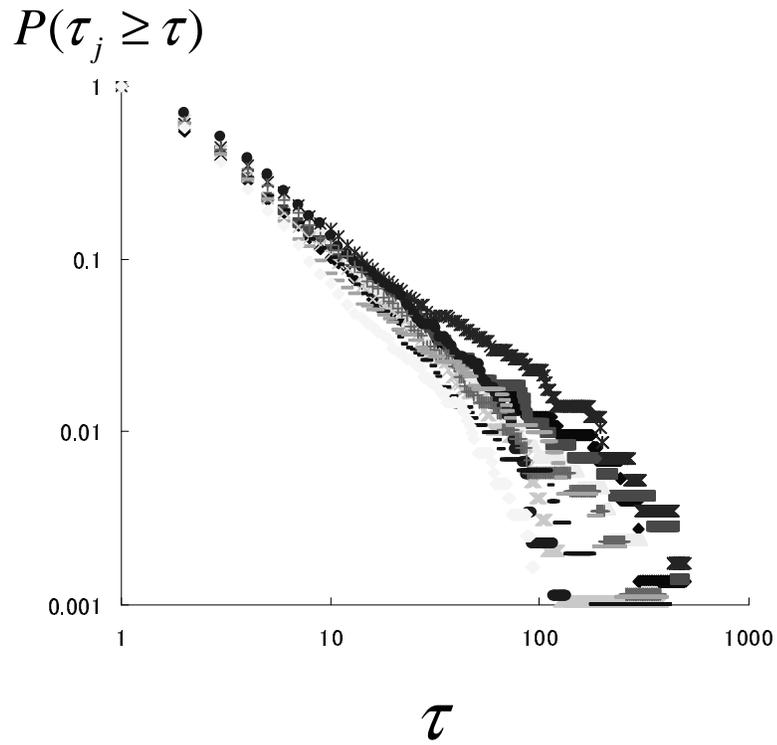}
\end{center}
\caption{The log-log plots of the complementary cumulative distributions of the calm-time intervals for ten companies with a threshold of $ \theta = 0.7 $. The ten companies were selected randomly from the 800 companies listed in the Tokyo Stock Exchange.} 
\label{fig3}
\end{figure}

\begin{figure}
\begin{center}
  \includegraphics[height=21cm]{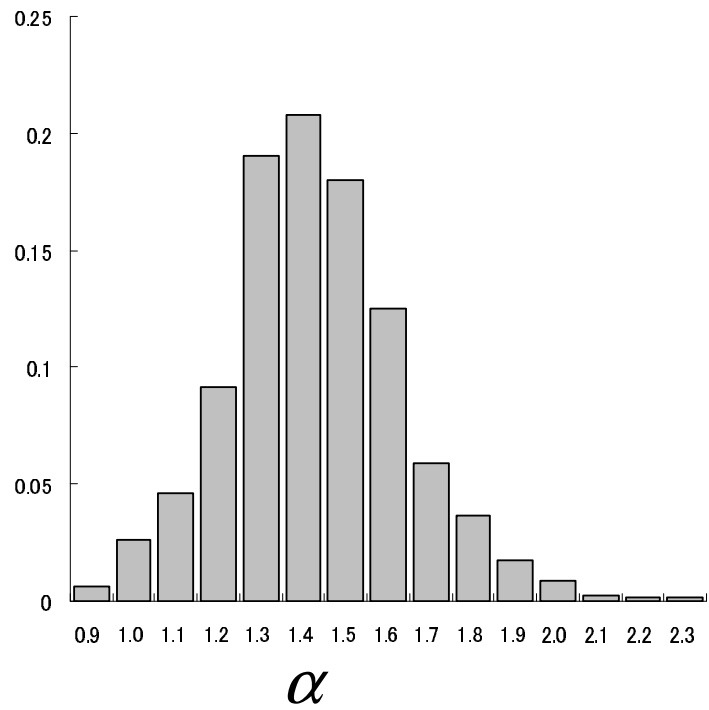}
\end{center}
\caption{The histogram of the power-law exponent $ \alpha $ estimated from the complementary cumulative distributions of the calm-time intervals for 800 companies listed in the Tokyo Stock Exchange with a threshold of $ \theta=0.7 $.} 
\label{fig4}
\end{figure}

\begin{figure}
\begin{center}
  \includegraphics[height=20cm,width=14cm]{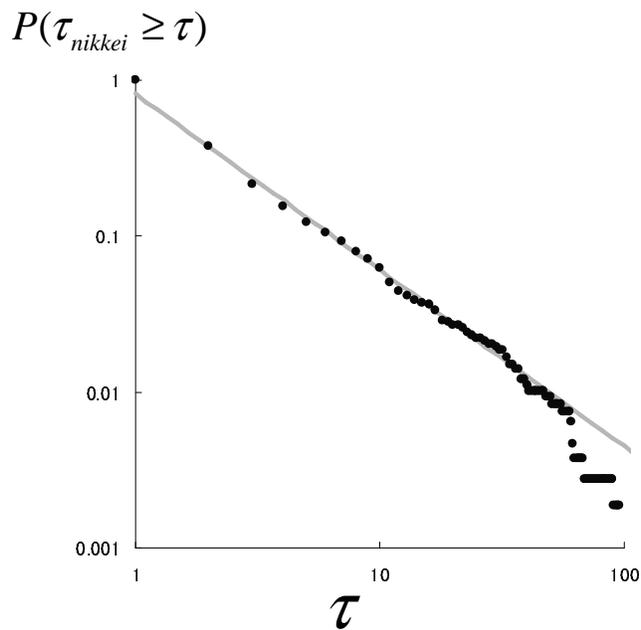}
\end{center}
\caption{The log-log plots of the survival function of the calm-time intervals for the Nikkei 225 index in the 27-year period from January 4, 1975 to December 28, 2001 with a threshold of $ \theta = 0.25 $. Dots represent the observed complementary cumulative distribution, and the solid lines represent the power-law function, $ P(\tau_{nikkei} \geq \tau) \sim \tau^{-\alpha} $ with the exponent $ \alpha = 1.13 $. } 
\label{fig5}
\end{figure}

\begin{figure}
\begin{center}
  \includegraphics[height=21cm,width=14cm]{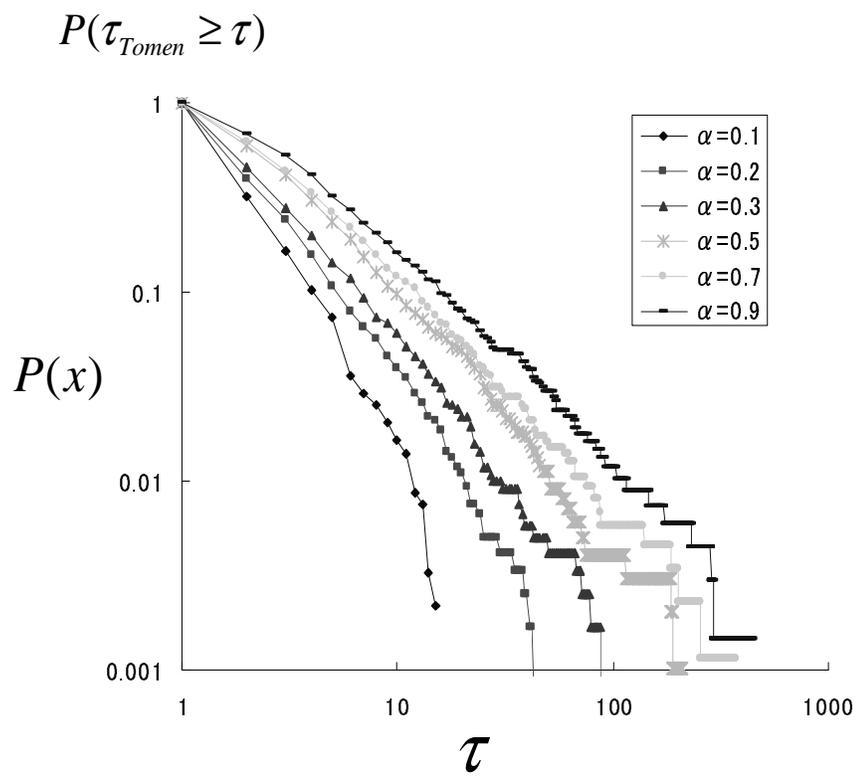}
\end{center}
\caption{The plots of the complementary cumulative distributions of the calm-time intervals for Tomen Corp. under the threshold $ \theta $ ranging from $0.1$ to $0.9$.} 
\label{fig6}
\end{figure}

\begin{figure}
\begin{center}
  \includegraphics[height=21cm,width=14cm]{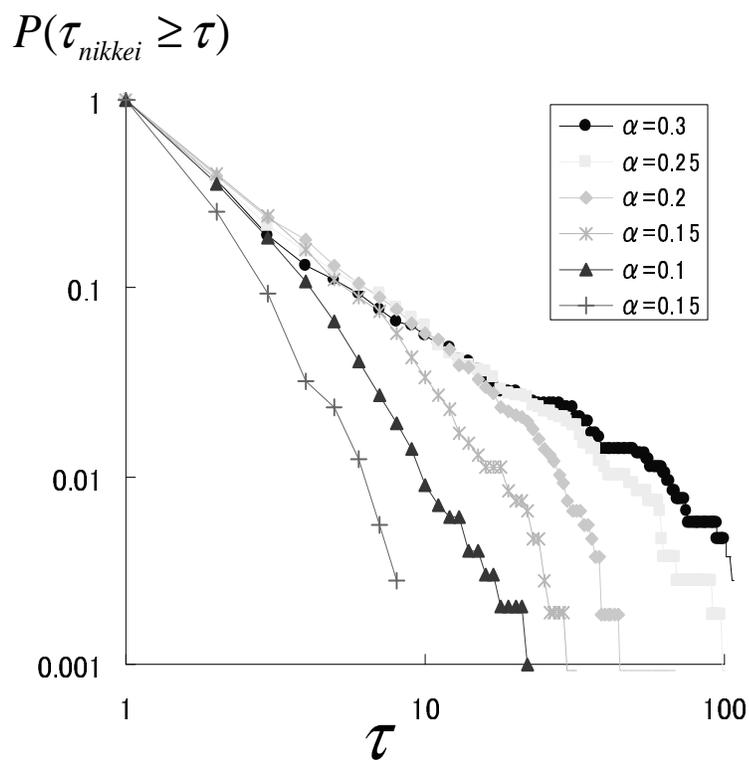}
\end{center}
\caption{The plots of the complementary cumulative distributions of the calm-time intervals for the Nikkei 225 index under the threshold $ \theta $ ranging from $0.15$ to $0.3$.} 
\label{fig7}
\end{figure}
\end{document}